\documentclass[5p,times,twocolumn]{elsarticle}

\usepackage{amsmath,amssymb}
\biboptions{comma,sort&compress}
\usepackage[utf8]{inputenc}
\usepackage{graphicx}
\usepackage{bm}
\usepackage{hyperref}
\hypersetup{
colorlinks = true,
linkcolor = blue,
anchorcolor = blue,
citecolor = blue,
filecolor = blue,
urlcolor = blue
}
\usepackage[dvipsnames]{xcolor}

\journal{Physics Letters B}

\begin{document}
\begin{frontmatter}

\title{Sequential Bayesian inference with correlated heavy-ion datasets}

\author[first,second]{Lipei Du}
\ead{ldu2@lbl.gov}
\affiliation[first]{organization={Department of Physics, University of California},
            city={Berkeley},
            postcode={94720}, 
            state={CA},
            country={USA}}
\affiliation[second]{organization={Nuclear Science Division, Lawrence Berkeley National Laboratory},
            city={Berkeley},
            postcode={94720}, 
            state={CA},
            country={USA}}
\date{\today}

\begin{abstract}
Bayesian inference provides a natural framework for updating knowledge as new
information becomes available, often in a sequential manner by incorporating
datasets in stages or reusing previous posteriors as priors.
A common implementation uses a factorized update in which datasets are
treated as conditionally independent. When datasets have statistically
correlated uncertainties, however, this approximation generally becomes
inconsistent with the joint likelihood and can lead to distorted posterior
constraints.
In this work, we investigate this issue in a controlled setting using
pseudo-data with a tunable covariance structure. We compare joint inference,
factorized sequential updating, and a formulation based on the exact
conditional likelihood.
We show that factorized updates reproduce the joint posterior under
conditional independence, but develop systematic deviations when correlations
are neglected, with discrepancies that increase with correlation strength
in our numerical examples.
In contrast, conditional sequential updating reproduces the joint
posterior exactly.
To interpret these deviations, we introduce an information decomposition
that separates the joint information into contributions from the first
dataset and the conditional information provided by the second.
By comparing marginal and conditional information, we show that correlations
induce structured changes in the constraints on different parameter
combinations, reflecting the interplay between covariance structure and
dataset sensitivities. This information-based analysis provides a quantitative
interpretation of the observed posterior discrepancies.
These results provide a practical diagnostic for assessing the consistency of
sequential Bayesian inference with correlated datasets and highlight the need
for a consistent treatment of correlations within a common probabilistic
framework.
\end{abstract}


\end{frontmatter}

\section{Introduction}\label{sec:introduction}

Bayesian inference has become a central framework for quantitative parameter
estimation in complex physical systems~\cite{Trotta:2008qt, MacKay:2003, Liddle:2009xe}. In relativistic heavy-ion collisions,
it is widely used to extract properties of the quark--gluon plasma by
confronting multi-observable datasets with dynamical models~\cite{Pratt:2015zsa,Bernhard:2016tnd, Bernhard:2019bmu, Nijs:2020ors, Nijs:2020roc,JETSCAPE:2020shq,JETSCAPE:2020mzn, JETSCAPE:2023ikg,JETSCAPE:2021ehl,Ehlers:2024miy,JETSCAPE:2026hdw,Du:2025rgq}. More broadly,
Bayesian methods provide a systematic way to update prior knowledge as new
information becomes available, making them a natural language for cumulative
scientific inference~\cite{Sivia:2006, vonToussaint:2011, Paquet:2023rfd}.

In many applications, information is acquired in stages. New datasets may
become available over time, or previous analyses may provide posterior
distributions that serve as inputs to subsequent studies~\cite{Bishop:2006, van2021bayesian}. This motivates the
idea of \emph{sequential Bayesian inference}, in which knowledge is progressively
refined as additional data are incorporated. Such an approach is attractive in
large-scale calibration problems, as it enables modular combination of
heterogeneous constraints and facilitates the propagation of prior information
across different stages of analysis.

A key question in this context is whether sequential updates remain consistent
with joint inference when multiple datasets are combined. In practice, such
updates are often carried out by incorporating datasets in stages or by
reusing posterior distributions from previous analyses as effective priors.
A common simplification is to assume that the datasets are conditionally
independent given the model parameters, so that the likelihood factorizes into
a product of marginal terms~\cite{Roch:2025jpu}.
While this assumption simplifies the analysis, it is not always made
explicit, and its validity requires the absence of residual statistical
dependence between datasets at fixed model parameters.

In realistic applications, conditional independence cannot always be
assumed a priori. Statistical correlations between datasets may arise from
shared experimental systematics, overlapping event samples, or correlated
theoretical uncertainties~\cite{JETSCAPE:2021ehl,Ehlers:2024miy,JETSCAPE:2026hdw}. In heavy-ion collisions, for example, particle yields,
transverse-momentum spectra, and anisotropic flow coefficients are all
sensitive to common features of the initial state and medium response~\cite{Bozek:2016yoj, Luzum:2012wu, Ollitrault:1992bk, Du:2023gnv,Teaney:2010vd,Du:2025tdh}.
More generally, different measurements often probe overlapping aspects of
a physical system~\cite{Moreland:2018gsh,Du:2025ith,Sorensen:2023zkk,Du:2025dot,Arslandok:2023utm, Du:2024wjm}.

When such correlations are present, the likelihood no longer factorizes, and
sequential updates based on marginal likelihoods do not, in general, reproduce
the result of a joint analysis. This leads to a fundamental inconsistency:
factorized sequential inference can yield posterior distributions that differ
systematically from those obtained from the full joint likelihood. While
effort has been devoted to representing and propagating complex
prior information across inference stages~\cite{Yamauchi:2023xrz,Roch:2025jpu},
the role of statistical correlations between datasets in sequential inference
has received comparatively less systematic attention. This is particularly
relevant since, even in joint analyses, cross-dataset correlations can be
challenging to fully quantify and are often treated approximately in practice~\cite{JETSCAPE:2021ehl,Ehlers:2024miy,JETSCAPE:2026hdw,Soltz:2024gkm},
so that their impact may already be nontrivial at the level of the full
likelihood.

The purpose of this work is to clarify the role of
statistical correlations in sequential Bayesian inference within a fixed
forward model. Rather than focusing
on a specific phenomenological application, we construct a reduced model that
captures the essential statistical structure of multi-observable inference
problems, and generate pseudo-data with a tunable covariance matrix that allows
the strength of cross-dataset correlations to be varied continuously.

Within this framework, we compare joint inference using the full likelihood,
factorized sequential updating, and a formulation based on the exact conditional
likelihood.
This setup enables a controlled assessment of how cross-dataset
correlations affect posterior inference.
In our numerical examples, factorized
updates reproduce the joint posterior when the datasets are conditionally
independent, but lead to systematic deviations when correlations are neglected,
with discrepancies that become more pronounced as the correlation strength
increases.
In contrast, conditional sequential updating reproduces the joint
posterior exactly and is independent of update order.
This equivalence follows from the general probability chain rule, while
the Gaussian setting allows the conditional likelihood to be constructed
analytically.

Beyond this comparison, we introduce a quantitative framework to analyze how
correlations redistribute information between datasets.
By comparing marginal and conditional information, we quantify how
correlations modify the additional information contributed by a dataset
after another has already been incorporated. We show that these effects
depend on the parameter combinations being constrained, reflecting the
interplay between covariance structure and dataset sensitivities.
This provides a direct link between the data covariance, the geometry of
parameter constraints, and the resulting deviations in posterior inference.

The remainder of this paper is organized as follows. In
Section~\ref{sec:model}, we introduce the reduced model and pseudo-data
construction. In Section~\ref{sec:bayes}, we review the Bayesian inference
framework and formulate the different sequential updating strategies.
In Section~\ref{sec:results}, we analyze the sensitivity structure, develop an
information-theoretic decomposition of correlated datasets, examine the
resulting posterior distortions, and quantify their magnitude using the
Kullback--Leibler divergence. Finally, Section~\ref{sec:summary} summarizes the
main findings and discusses their broader implications.

\section{Reduced model and pseudo-data setup}\label{sec:model}

To study the role of correlations in a controlled setting, we consider a reduced
model that captures the essential structure of multi-observable inference
problems \cite{Moreland:2018gsh,Du:2025rgq,JETSCAPE:2020mzn}. The model is designed to isolate, in a minimal and transparent way, how different classes of observables constrain
overlapping parameter directions. Details of the construction are provided in
\ref{app:model}.

We introduce four model parameters,
\begin{equation}
    \theta = (S,\, \varepsilon_2,\, \kappa_N,\, \kappa_v),
\end{equation}
where $S$ denotes an entropy-like scale, $\varepsilon_2$ an eccentricity-like
parameter, and $\kappa_N$ and $\kappa_v$ characterize the response of
yield-like and flow-like observables, respectively. The observable vector
consists of two groups,
\begin{equation}
d = (N_1, N_2, N_3, v_{2,1}, v_{2,2}, v_{2,3}),
\end{equation}
corresponding to yield-like quantities $N_i$ and flow-like coefficients
$v_{2,i}$ evaluated in three representative centrality bins.

The model $m(\theta)$ encodes a simple mapping between parameters and observables. The
dominant dependence is captured by sector-wide amplitudes, with the yield-like
observables $N_i$ primarily controlled by the combination $\kappa_N S$ and the
flow-like observables $v_{2,i}$ by $\kappa_v \varepsilon_2$. Weak
cross-sensitivity between the two sectors is introduced through small mixing
terms, ensuring that each observable group retains a subleading dependence on
the parameters associated with the other sector. This structure is motivated by
realistic heavy-ion simulations \cite{Heinz:2013th}, where particle yields are
largely governed by the overall entropy scale~\cite{Giacalone:2019ldn,Du:2022yok,Du:2023efk}, while anisotropic flow reflects the response to the initial-state eccentricity~\cite{Alver:2010gr, Gardim:2011xv}.
These shared parameter sensitivities are distinct from the statistical
correlations between observable groups, which we introduce below through
the pseudo-data covariance.

The centrality dependence is encoded through reference values
$N_i^{\mathrm{ref}}$ and $v_{2,i}^{\mathrm{ref}}$,
which are guided by representative experimental measurements in different
centrality classes \cite{ALICE2010, ALICE:2011ab} and serve to anchor the model to realistic observable scales,
reflecting the variation of entropy and eccentricity across these bins. The model predictions are constructed by rescaling these reference
values through parameter-dependent amplitudes, supplemented by mild bin-to-bin
modulations so that different centrality bins probe slightly different
directions in parameter space.

\begin{figure}[!t]
\centering
\includegraphics[width=\linewidth]{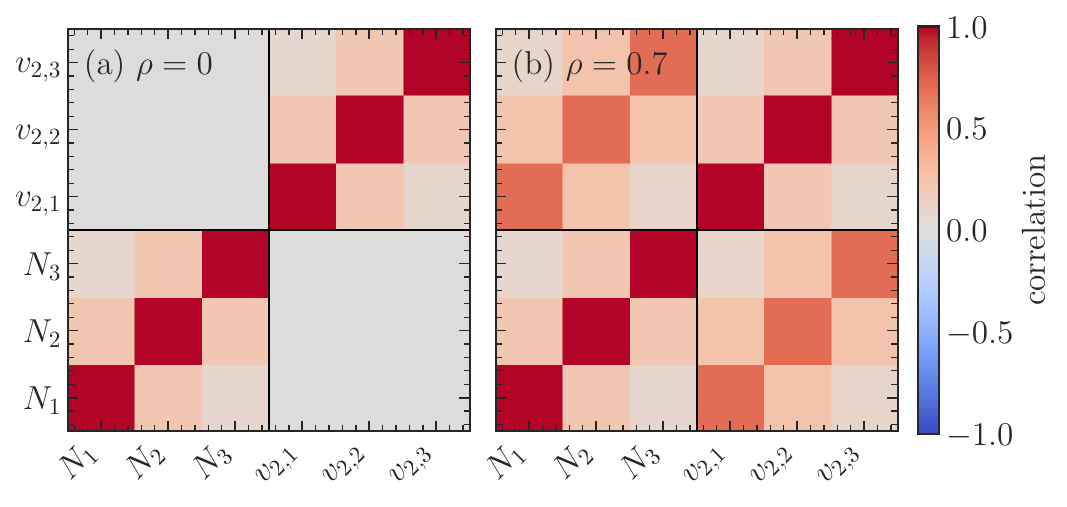}
\caption{
Correlation matrices for the pseudo-data at two representative values of the
cross-dataset correlation strength $\rho$. The observable ordering is
$(N_1, N_2, N_3, v_{2,1}, v_{2,2}, v_{2,3})$, with guide lines separating the
two observable groups.
}
\label{fig1}
\end{figure}

An important advantage of this setup is that the forward model is analytic and
inexpensive to evaluate, allowing posterior distributions to be computed
directly without the need for surrogate models or emulators
\cite{Paquet:2023rfd,Rasmussen:2005}. This avoids additional sources of
approximation and uncertainty that are unrelated to the statistical questions
addressed here, and ensures that any differences observed between inference strategies can be
unambiguously attributed to their statistical treatment, rather than to model
approximations.
We adopt broad uniform priors over all parameters, so that the posterior is primarily driven by the likelihood.

Pseudo-data are generated from a fixed reference parameter point $\theta_\star$
by adding Gaussian fluctuations,
\begin{equation}
d = m(\theta_\star) + \delta, \qquad \delta \sim \mathcal{N}(0,\Sigma),
\end{equation}
where $m(\theta)$ denotes the model prediction and $\Sigma$ is the covariance
matrix. The covariance is written in block form as
\begin{equation}\label{eq:covariance}
\Sigma =
\begin{pmatrix}
\Sigma_1 & \Sigma_{12} \\
\Sigma_{21} & \Sigma_2
\end{pmatrix},
\end{equation}
where $\Sigma_1$ and $\Sigma_2$ describe correlations within the yield and
flow sectors, and $\Sigma_{12}$ encodes correlations between them. Within each
sector, centrality bins are correlated over a finite range, introducing
nontrivial bin-to-bin correlations while retaining a component of
independent fluctuations.

Cross-dataset correlations are introduced through the off-diagonal block
$\Sigma_{12} = \rho\, \tilde{\Sigma}_{12}$ \cite{Soltz:2024gkm}, where
$\tilde{\Sigma}_{12}$ defines a fixed correlation pattern and the parameter
$\rho$ controls the overall strength of correlations between the two observable
groups.
The covariance construction also permits negative values of $\rho$,
corresponding to anti-correlations, provided that the full covariance matrix
remains positive definite.
The resulting correlation matrices are illustrated in Fig.~\ref{fig1}.
The limit $\rho=0$ corresponds to statistically independent datasets,
while positive values introduce cross-dataset correlations of increasing
strength.
Our numerical study uses $0\leq\rho\leq0.7$, within the
positive-definite range of the covariance matrix.
For each value of $\rho$, pseudo-data are generated from
$\mathcal{N}[m(\theta_\star),\Sigma(\rho)]$ using the same random seed.
All inference strategies use the same pseudo-data realization at fixed
$\rho$, while the realized data vector generally changes with $\rho$.
By varying $\rho$, we examine the impact of different covariance
configurations and their associated pseudo-data realizations on Bayesian
inference.
In the following, $\rho$ serves as the primary control parameter of
the study.

\section{Sequential Bayesian inference framework}\label{sec:bayes}

With this setup in place, we now turn to the statistical framework for
inferring model parameters $\theta$ from two datasets, $D_1$ and $D_2$.
In our demonstration, these correspond to the two observable groups
introduced in Section~\ref{sec:model}, namely yield-like observables
$D_1 = \{N_i\}$ and flow-like observables $D_2 = \{v_{2,i}\}$, although the
discussion below is general and does not depend on their specific physical
interpretation. The model prediction for the full dataset is written as
$m(\theta)$, and the measured data vector is denoted by $d$.
The joint posterior distribution is given by Bayes' theorem \cite{vonToussaint:2011,Sivia:2006},
\begin{equation}\label{eq:full}
p(\theta|D_1,D_2)
\propto
p(D_1,D_2|\theta)\, p(\theta),
\end{equation}
where $p(\theta)$ is the prior and $p(D_1,D_2|\theta)$ is the likelihood for the
combined dataset. This expression defines the reference result against which all
other inference procedures will be compared.

Using the chain rule of probability, the joint likelihood can be written exactly as
$p(D_1,D_2|\theta)
=
p(D_2|D_1,\theta)\, p(D_1|\theta)$,
which leads to the sequential form of the posterior,
\begin{equation}\label{eq:seq}
p(\theta|D_1,D_2)
\propto
p(D_2|D_1,\theta)\, p(\theta|D_1),
\end{equation}
where $p(\theta|D_1) \propto p(D_1|\theta)\,p(\theta)$ denotes the posterior
obtained from the first dataset alone.
This relation shows that sequential Bayesian updating is, in principle, always
well-defined: once the posterior $p(\theta|D_1)$ is obtained, the second update
is governed by the conditional likelihood $p(D_2|D_1,\theta)$. The question is
therefore not whether sequential inference is valid, but how this conditional
dependence is treated in practice.

A commonly used approximation is to replace the conditional likelihood by the
marginal likelihood $p(D_2|\theta)$, leading to a factorized sequential
update,
\begin{equation}\label{eq:fact}
p_{\mathrm{fact}}(\theta|D_1,D_2)
\propto
p(D_2|\theta)\, p(\theta|D_1).
\end{equation}
This form is widely adopted in multi-stage Bayesian analyses, where posterior
distributions from earlier datasets are used as effective priors for subsequent
updates. Under the assumption that the datasets are conditionally independent,
$p(D_2|D_1,\theta) = p(D_2|\theta)$,
the factorized update reproduces the joint posterior exactly.

When the datasets are correlated, the assumption of conditional independence
is violated, and the conditional and marginal likelihoods
generally differ,
$p(D_2|D_1,\theta) \neq p(D_2|\theta)$. In this case, the factorized update
does not, in general, reproduce
the result of a joint analysis.
The discrepancy arises because the factorized likelihood neglects the
statistical dependence between datasets, treating their fluctuations as
conditionally independent. In contrast, the conditional likelihood
$p(D_2|D_1,\theta)$ accounts for the correlations with $D_1$ when incorporating
the additional constraints from $D_2$. Ignoring these correlations can
therefore distort the posterior, including its location, width, and
correlations between parameters. Depending on the covariance structure and
parameter sensitivities, the factorized treatment may either overestimate
or underestimate the information contributed by the second dataset.
In this sense, the factorized sequential update corresponds to an inference in
which cross-dataset correlations are neglected. The resulting discrepancy
therefore reflects a more general consequence of ignoring correlations in the
likelihood, independent of whether the analysis is performed sequentially or
jointly.

These considerations can be made explicit for Gaussian likelihoods. We consider
a Gaussian likelihood for the full data vector $d = (d_1, d_2)$,
\begin{equation}\label{eq:gaussian_joint}
\log \mathcal{L}_{\mathrm{joint}}(\theta)
=
-\frac12 \big(d - m(\theta)\big)^T \Sigma^{-1} \big(d - m(\theta)\big)
-\frac12 \log \det \Sigma + \mathrm{const},
\end{equation}
where $\Sigma$ is the covariance matrix \eqref{eq:covariance} defined in
Section~\ref{sec:model}.
In this study, $\Sigma$ describes the prescribed pseudo-data covariance;
additional theoretical-model or emulator uncertainties are not included.
The conditional structure of the likelihood can be made explicit by
reorganizing the quadratic form in Eq.~\eqref{eq:gaussian_joint}, separating contributions associated
with $d_1$ and $d_2$. This decomposition, derived
explicitly in \ref{app:matrix}, leads to a factorized representation in
which the second term corresponds to the conditional likelihood
$p(D_2|D_1,\theta)$.

For a Gaussian likelihood, the conditional distribution $p(D_2|D_1,\theta)$
takes a Gaussian form, with covariance given by the Schur complement,
\begin{equation}\label{eq:schur}
\Sigma_{2|1} = \Sigma_2 - \Sigma_{21}\Sigma_1^{-1}\Sigma_{12},
\end{equation}
and mean
\begin{equation}\label{eq:mean}
\mu_{2|1}(\theta)
=
m_2(\theta)
+
\Sigma_{21}\Sigma_1^{-1}
\bigl(d_1 - m_1(\theta)\bigr),
\end{equation}
where $m_1(\theta)$ and $m_2(\theta)$ denote the model predictions restricted
to the two datasets.\footnote{%
The expressions above coincide with the standard conditioning formulas of a
multivariate Gaussian distribution. In Gaussian process regression and emulator
construction, the same relations define predictive means and covariances
conditioned on training data \cite{Rasmussen:2005}, with the training set playing
a role analogous to $D_1$.
}
These expressions follow directly from the block structure of the covariance
matrix and its inverse, as detailed in \ref{app:matrix}.
The corresponding conditional log-likelihood $\log p(D_2|D_1,\theta)$ is
\begin{equation}
\begin{aligned}
\log \mathcal{L}_{2|1}(\theta)
={}&-\frac12 \big(d_2-\mu_{2|1}(\theta)\big)^T
\Sigma_{2|1}^{-1}
\big(d_2-\mu_{2|1}(\theta)\big)\\
&-\frac12\log\det\Sigma_{2|1}
+\mathrm{const}.
\end{aligned}
\end{equation}
The conditional mean \eqref{eq:mean} accounts for the component of the
fluctuations in $d_2$ that can be statistically predicted from the residuals
$d_1-m_1(\theta)$, while the Schur-complement covariance \eqref{eq:schur}
describes the uncertainty remaining after conditioning on $d_1$.
The resulting conditional residual retains its dependence on the model
parameters and can therefore provide additional constraints.
Together, these terms ensure that the update
correctly accounts for the statistical dependence between datasets,
thereby reproducing the joint likelihood.

Importantly, strong statistical correlations do not necessarily imply
that the second dataset contains little additional information about the
model parameters. The conditional likelihood depends not only on the
remaining statistical uncertainty but also on how the model predictions
for the two datasets respond to changes in the parameters. Correlations
can therefore either enhance or reduce constraints along different
parameter directions, depending on the interplay between the covariance
structure and parameter sensitivities. This distinction provides the
basis for the information decomposition developed in the following
section.

\section{Results and discussion}\label{sec:results}

We examine how correlations between datasets affect sequential Bayesian
inference within the controlled setup introduced above. To this end, we
compare three inference strategies: joint inference using the full likelihood
[Eq.~\eqref{eq:full}], sequential updating with a factorized likelihood
[Eq.~\eqref{eq:fact}], and sequential updating using the conditional
likelihood $p(D_2|D_1,\theta)$ [Eq.~\eqref{eq:seq}]. This comparison isolates
the role of cross-dataset correlations and provides a direct test of how
different implementations of sequential inference impact the resulting
posterior. The results are organized to progressively connect:
(i) the structure of parameter constraints (Section~\ref{subsec1}),
(ii) the decomposition of information between datasets (Section~\ref{subsec2}),
(iii) the resulting posterior distortions (Section~\ref{subsec3}),
(iv) their quantitative impact (Section~\ref{subsec4}), and
(v) their broader interpretation and practical applications
(Section~\ref{subsec5}).

For a Gaussian likelihood with parameter-independent covariance, the
Fisher information matrix characterizes the local parameter sensitivities of
the likelihood. Writing $J=\partial m/\partial\theta$ for the model Jacobian,
it takes the form $F = J^T \Sigma^{-1} J$. As shown in \ref{app:matrix},
this structure follows directly from the quadratic form of the Gaussian
likelihood and makes explicit how cross-dataset correlations enter parameter
inference through couplings between different observables. In the following,
we use the Fisher matrix as a local measure of information to analyze how
correlations modify the contribution of different datasets.

\subsection{Sensitivity structure and parameter constraints}\label{subsec1}

We first examine the parameter sensitivities of the two observable
groups. As distinguished above, these shared sensitivities are separate from
the statistical cross-dataset correlations encoded in the covariance matrix.
Together, they determine how the correlations affect different directions in
parameter space. To this end, we introduce a weighted sensitivity matrix
\begin{equation}
\widetilde J_{\alpha i}
=
\frac{1}{\sigma_\alpha}
\frac{\partial m_\alpha}{\partial \theta_i},
\end{equation}
which measures the local response of each observable in units of its statistical
uncertainty. The sensitivity matrix also provides a direct link to the
information content of the likelihood in parameter space. In terms of the
Fisher information matrix $F = J^T \Sigma^{-1} J$, for uncorrelated observables,
where the covariance is diagonal,
$\Sigma_{\alpha\beta} = \sigma_\alpha^2 \delta_{\alpha\beta}$, this reduces to
$F_{ij} = \sum_{\alpha} \widetilde J_{\alpha i}\widetilde J_{\alpha j}$.
In this sense, the weighted sensitivities provide a simple diagnostic of
observable responses, although the full Fisher matrix is needed to characterize
constraints on correlated parameter combinations.

\begin{figure}[!t]
\centering
\includegraphics[width=0.7\linewidth]{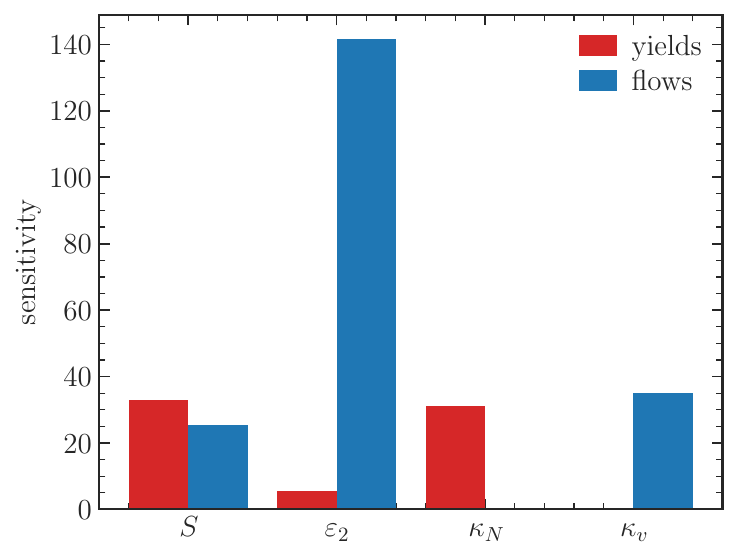}
\caption{
Aggregated sensitivity strength of the yield-like ($D_1$) and flow-like ($D_2$)
observable groups to each model parameter, evaluated at the reference point $\theta_\star$.
}
\label{fig2}
\end{figure}

For each dataset $g \in \{D_1, D_2\}$, we define an aggregated sensitivity strength
\begin{equation}
S_i^{(g)} =
\left[
\sum_{\alpha \in g}
{\widetilde J_{\alpha i}}^2
\right]^{1/2},
\end{equation}
which provides a compact measure of the local sensitivity of dataset $g$
to parameter $\theta_i$, rather than a marginalized parameter constraint.
The resulting sensitivity structure, shown in
Fig.~\ref{fig2}, exhibits a clear hierarchy. The yield-like observables
primarily constrain $S$ and $\kappa_N$, while the flow-like observables
primarily constrain $\varepsilon_2$ and $\kappa_v$. In addition, the
cross-sector sensitivity is asymmetric: the flow observables retain a
non-negligible dependence on $S$, whereas the yield observables are only weakly
sensitive to $\varepsilon_2$. This asymmetry follows directly from the model
construction\footnote{%
In realistic heavy-ion dynamics \cite{Heinz:2013th}, a larger entropy scale $S$ generally leads to
a longer system lifetime and enhanced development of anisotropic flow, resulting
in larger $v_2$. In contrast, variations in the initial eccentricity
$\varepsilon_2$ have only a weak impact on integrated yields. This motivates the
asymmetric cross-sector sensitivity in the reduced model.
}
and provides a minimal realization of partially overlapping constraints.

\subsection{Information decomposition and cross-dataset correlations}\label{subsec2}

These sensitivity patterns identify the parameter directions probed by both
datasets, and hence where cross-dataset correlations may influence inference.
We now quantify this effect by examining how the covariance structure modifies
the information content of the likelihood. As shown in \ref{app:matrix},
this modification can be understood in terms of the conditional decomposition
of the likelihood, which induces a corresponding decomposition of Fisher
information.

We focus on the contribution of the second dataset $D_2$. When correlations are
ignored, its contribution to parameter constraints is described by the Fisher
matrix
\begin{equation}
F_2 = J_2^T \Sigma_2^{-1} J_2,
\end{equation}
where $J_2 = \partial m_2 / \partial \theta$ is the Jacobian of the model
prediction for $D_2$. This matrix represents the information that $D_2$
appears to provide when treated as statistically independent from $D_1$.

When the datasets are correlated, the information contributed by $D_2$
after incorporating $D_1$ generally differs from its marginal contribution.
The relevant quantity for sequential inference is therefore the conditional
contribution obtained from $p(D_2|D_1,\theta)$, which accounts for the
statistical dependence between datasets. For a
Gaussian likelihood, this distribution has covariance $\Sigma_{2|1}$ given by
the Schur complement [Eq.~\eqref{eq:schur}] and mean
$\mu_{2|1}(\theta)$ [Eq.~\eqref{eq:mean}]. The corresponding Jacobian follows
from differentiating the conditional mean,
\begin{equation}
J_{2|1}
=
\frac{\partial \mu_{2|1}(\theta)}{\partial \theta}
=
J_2 - \Sigma_{21}\Sigma_1^{-1} J_1,
\end{equation}
as derived explicitly in \ref{app:matrix},
showing that conditioning modifies the parameter response of $D_2$ by
subtracting a covariance-weighted contribution from $D_1$. This modification
need not reduce the sensitivity along every parameter direction.
The conditional Fisher matrix then becomes
\begin{equation}
F_{2|1}
=
J_{2|1}^T \Sigma_{2|1}^{-1} J_{2|1},
\end{equation}
which quantifies the additional local information about the parameters
provided by $D_2$ after $D_1$ has been incorporated. For the Gaussian
likelihood with parameter-independent covariance, the joint Fisher information
decomposes exactly as $F_{\mathrm{joint}}=F_1+F_{2|1}$, as shown in
\ref{app:matrix}.

\begin{figure}[!t]
\centering
\includegraphics[width=\linewidth]{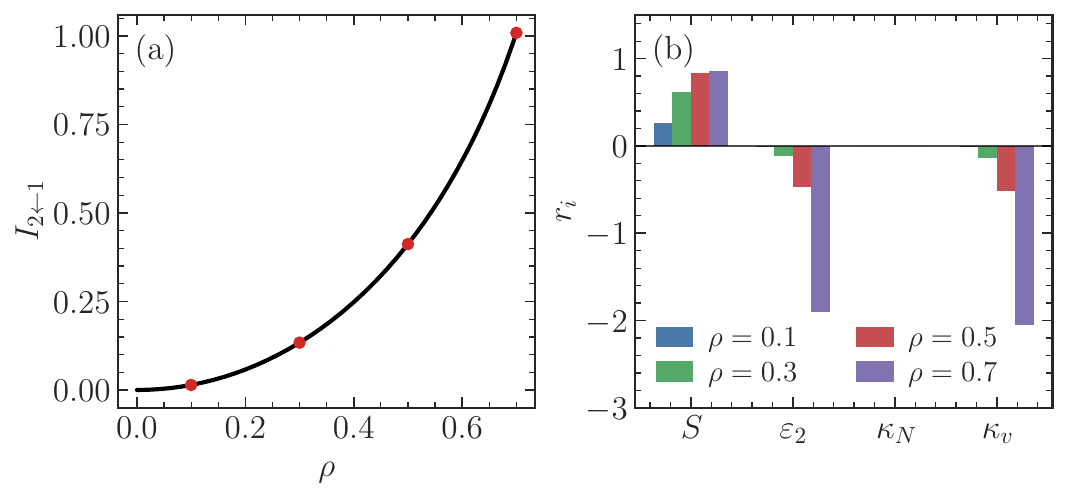}
\caption{
(a) Data-space information flow $I_{2\leftarrow 1}$ as a function of the correlation strength $\rho$.
(b) Corresponding signed parameter-resolved information difference $r_i$ for selected values of $\rho$.
}
\label{fig3}
\end{figure}

This separation can also be understood directly at the level of the data,
where it is expressed in terms of the covariance matrices. The conditional
covariance $\Sigma_{2|1}$ represents the uncertainty in $D_2$ that remains after
accounting for correlations with $D_1$, while the difference between $\Sigma_2$
and $\Sigma_{2|1}$ encodes the component of $D_2$ that is predictable from
$D_1$. A convenient scalar measure of this effect, which we refer to as the
information flow from $D_1$ to $D_2$, is given by
\begin{equation}
I_{2\leftarrow 1}
=
\frac{1}{2}
\log \frac{\det \Sigma_2}{\det \Sigma_{2|1}},
\end{equation}
which quantifies the reduction in uncertainty volume of $D_2$ due to
knowledge of $D_1$. For a jointly Gaussian distribution, this is the mutual
information between $D_1$ and $D_2$ at fixed model parameters. It measures
data-space statistical dependence, rather than information about the model
parameters.

Conditioning also modifies the Fisher information in parameter space,
through both the conditional covariance and the conditional parameter response.
The difference $\Delta F_{2\leftarrow 1}=F_2-F_{2|1}$ therefore characterizes
the change between marginal and conditional information. This difference is
not necessarily positive semidefinite. To characterize its effect along the
individual parameter coordinates, we introduce the signed measure
\begin{equation}
r_i
=
1 - \frac{(F_{2|1})_{ii}}{(F_2)_{ii}},
\end{equation}
which measures the relative change in the diagonal Fisher contribution
associated with parameter $\theta_i$. Positive values indicate a reduction in
conditional information relative to $F_2$, whereas negative values indicate
an enhancement.\footnote{%
This definition is based on the diagonal elements of the Fisher matrix and
therefore depends on the chosen parameter basis. In general, the Fisher matrix
is not diagonal, and parameter constraints are determined by its full
eigenstructure. As a result, individual parameters do not correspond uniquely to
independent directions in parameter space, and correlations may act on linear
combinations of parameters rather than on individual components.
}

The behavior of these quantities is illustrated in Fig.~\ref{fig3}.
Figure~\ref{fig3}(a) shows that the information flow $I_{2\leftarrow 1}$ increases
with the correlation strength $\rho$, indicating increasing statistical
predictability of $D_2$ from $D_1$.
The corresponding change in parameter information is non-uniform because
it also depends on the conditional parameter response.
Figure~\ref{fig3}(b) shows that the signed measure $r_i$ varies strongly
across parameter coordinates. This pattern can be understood in connection with the sensitivity
structure shown in Fig.~\ref{fig2}. Parameters such as $\varepsilon_2$ and $\kappa_v$, which are primarily
constrained by the flow observables ($D_2$) but remain indirectly coupled to
$D_1$ through shared dependencies (notably via $S$), exhibit the largest
deviations. In contrast,
parameters such as $\kappa_N$, to which $D_2$ is only weakly sensitive, remain
largely unaffected, while parameters like $S$, which are constrained by both
datasets, show more moderate behavior.

For $\varepsilon_2$ and $\kappa_v$, $r_i$ becomes negative, indicating
that conditioning enhances the corresponding diagonal Fisher contributions
rather than reducing them. The dependence on $\rho$ also differs by parameter
coordinate: $S$ exhibits a gradual increase in $r_i$, while $\varepsilon_2$
and $\kappa_v$ decrease to large negative values. These differences reflect
the interplay between correlated fluctuations and the sensitivities of the
observables to different parameter combinations.

Together, these results show that the conditional decomposition separates
the information provided by $D_1$ from the additional information contributed
by $D_2$ after conditioning. Comparing the latter with the marginal information
of $D_2$ reveals a correlation-induced redistribution of information across
parameter combinations, rather than a universally positive redundancy.
We next examine how these differences appear in the posterior distributions.

\subsection{Impact on posterior inference}\label{subsec3}

Having established how correlations modify the information carried by $D_2$, we
now examine how these effects appear in the posterior distribution. As discussed
in Section~\ref{sec:bayes}, factorized sequential updating reproduces the joint
posterior under conditional independence, so deviations from the
joint result reveal the impact of neglected correlations in our comparison.

\begin{figure}[!t]
\centering
\includegraphics[width=\linewidth]{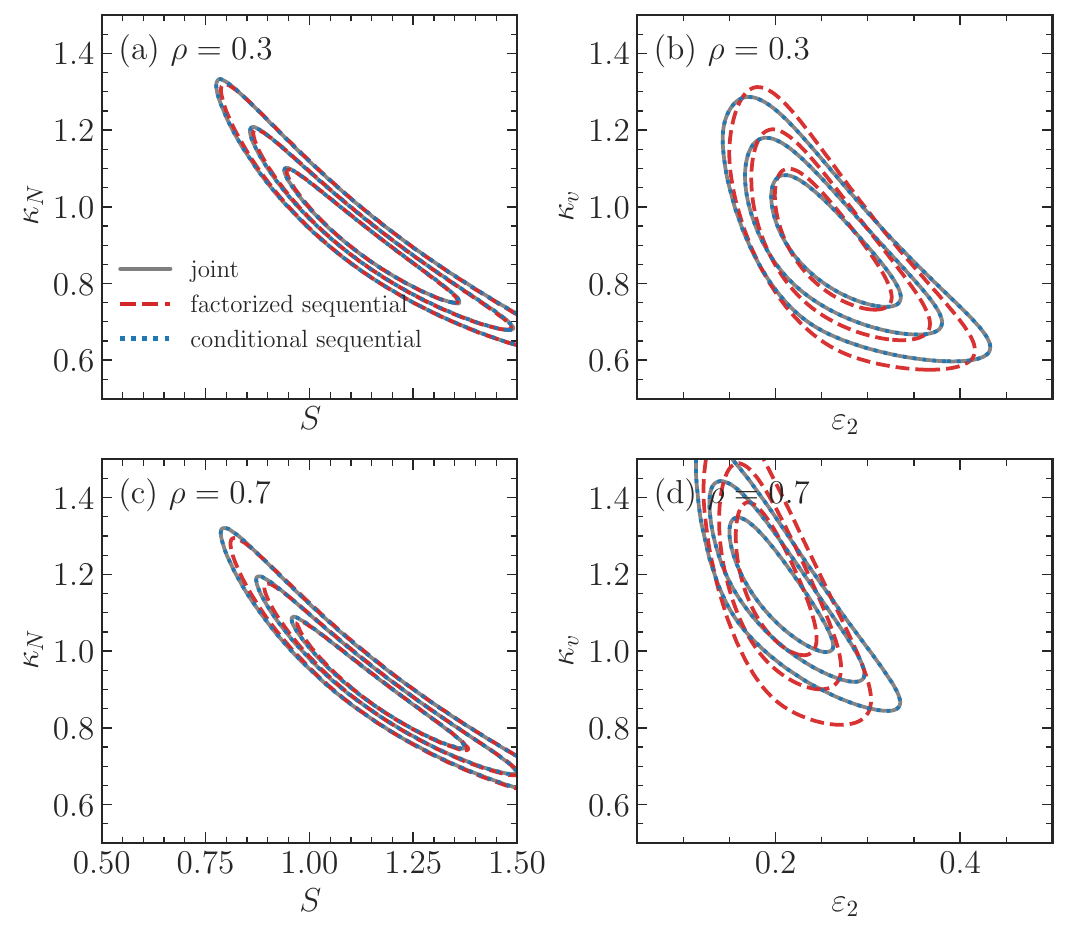}
\caption{
Posterior contours for joint inference (black solid), factorized sequential updating (red dashed), and conditional sequential updating (blue dotted), shown at $\rho=0.3$ (top) and $\rho=0.7$ (bottom). Left panels correspond to the $(S,\kappa_N)$ plane, and right panels to the $(\varepsilon_2,\kappa_v)$ plane.
}
\label{fig4}
\end{figure}

The effect is illustrated in Fig.~\ref{fig4}. For $\rho=0$, all inference
methods coincide (not shown), consistent with the factorization of the
likelihood. As $\rho$ increases, systematic deviations develop between the
factorized sequential result and the joint posterior, while the conditional
sequential result remains consistent with the joint inference. In this numerical
study, the magnitude of these deviations increases with $\rho$.

We have also verified that sequential updating based on the conditional
likelihood yields results that are independent of the order in which the
datasets are incorporated, in agreement with the joint posterior (not shown).
In contrast, factorized sequential updates are likewise order-independent,
since they correspond to a product of marginal likelihood terms, but generally
do not reproduce the joint result once cross-dataset correlations are present.

The structure of these deviations can be understood in terms of the
information decomposition introduced in Section~\ref{subsec2}. The factorized
sequential update neglects the distinction between marginal and conditional
information, leading to distortions along directions where this difference is
largest.
This is reflected in Fig.~\ref{fig4}, where the deviations are most
pronounced in the $(\varepsilon_2,\kappa_v)$ plane, while remaining
comparatively mild in the $(S,\kappa_N)$ plane. This pattern reflects how
correlations modify the effective information associated with different
parameter directions: directions strongly affected by correlations exhibit the
largest distortions, while those for which the marginal and conditional
structures remain similar are comparatively stable.

More generally, the posterior distortions reflect changes in the
conditional information across parameter directions, rather than simply the
removal of redundant information or a uniform overcounting of information.

\subsection{Quantifying deviations from joint inference}\label{subsec4}

Having established the qualitative behavior, we now quantify the deviation
between sequential inference and the joint result using the
Kullback--Leibler (KL) divergence \cite{kullback1951information} between a sequential posterior
$p_{\mathrm{seq}}(\theta)$ and the joint posterior
$p_{\mathrm{joint}}(\theta)$,
\begin{equation}
D_{\mathrm{KL}}(p_{\mathrm{seq}} \,\|\, p_{\mathrm{joint}})
=
\int d\theta\,
p_{\mathrm{seq}}(\theta)\,
\log \frac{p_{\mathrm{seq}}(\theta)}{p_{\mathrm{joint}}(\theta)}.
\end{equation}
This quantity measures the discrepancy between the sequential posterior
and the joint posterior, including differences in their locations and shapes.\footnote{%
KL divergence is commonly used in Bayesian analyses to quantify the information
gain from prior to posterior distributions. In heavy-ion physics, it has been
used, for example, to assess the constraining power of different observables
by the JETSCAPE Collaboration~\cite{JETSCAPE:2020shq,JETSCAPE:2020mzn,Putschke:2019yrg}.
}

\begin{figure}[!t]
\centering
\includegraphics[width=\linewidth]{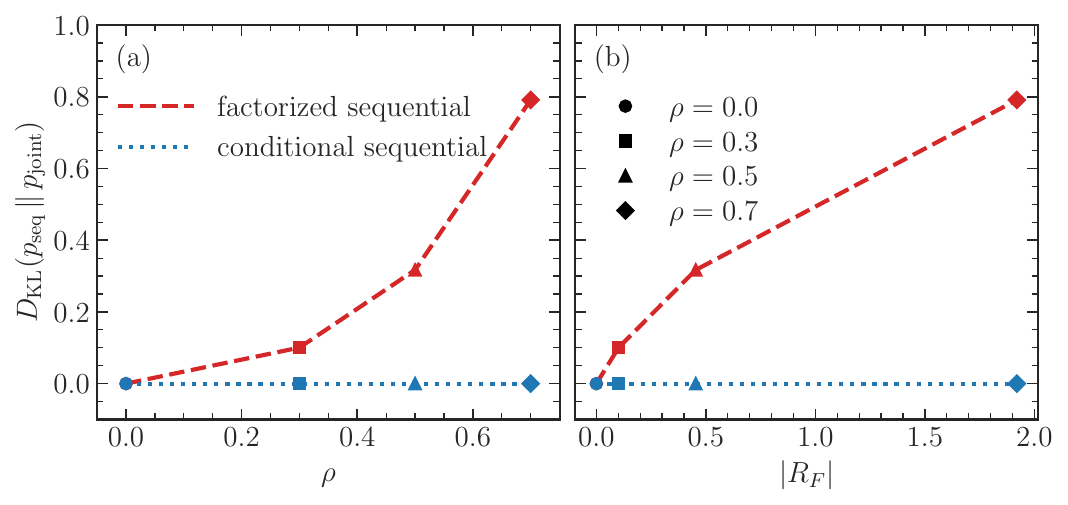}
\caption{
(a) Kullback--Leibler (KL) divergence between sequential posteriors and the
joint posterior as a function of the correlation strength $\rho$, for
factorized (red) and conditional (blue) sequential updating.
(b) KL divergence as a function of the magnitude of the normalized
Fisher-information difference, $|R_F|$.
}
\label{fig5}
\end{figure}

The results are shown in Fig.~\ref{fig5}. Panel (a) displays the KL divergence as
a function of the correlation strength $\rho$. For $\rho=0$, the divergence
vanishes for both sequential methods, reflecting the factorization of the
likelihood in the absence of cross-dataset correlations. In our numerical
study, the divergence for factorized sequential updating increases monotonically
with $\rho$, while the conditional sequential result remains consistent with
the joint posterior over the full range. Within this scan, the divergence
grows smoothly, with no apparent threshold behavior.

Panel (b) compares the KL divergence with the difference between marginal
and conditional Fisher information introduced in Section~\ref{subsec2}.
Specifically, we consider the normalized, signed Fisher-information difference
\begin{equation}
R_F =
\frac{\mathrm{Tr}(F_2 - F_{2|1})}{\mathrm{Tr}(F_2)}.
\end{equation}
This quantity summarizes the net change in Fisher information in the
chosen parameter coordinates; it can have either sign and is not invariant
under a rescaling or reparameterization of the model parameters.
For each covariance configuration, $R_F$ is computed from the underlying
model sensitivities and full covariance matrix, providing a common diagnostic
against which the posterior deviations from both sequential inference
methods are compared. In our numerical study, Fig.~\ref{fig5}(b) shows a
monotonic relation between $D_{\mathrm{KL}}$ and the magnitude $|R_F|$ for
the factorized sequential update, illustrating the association between the
marginal--conditional information mismatch and posterior deviations.
In contrast, the conditional sequential update reproduces the joint posterior,
yielding vanishing KL divergence. The quantity $R_F$ is therefore a useful
diagnostic within the present setup, rather than a universal predictor of
posterior discrepancies.

The KL divergence provides a complementary perspective to the parameter-level
analysis presented in Section~\ref{subsec3}. While Fig.~\ref{fig4} illustrates
posterior deviations in selected parameter subspaces, the KL divergence
quantifies the discrepancy between the full posterior distributions.
It therefore provides an overall measure of the difference between
inference methods, without specifying which parameter combinations
contribute most strongly to that difference.

\subsection{Interpretation and practical applications}\label{subsec5}

These results highlight a fundamental source of discrepancy in multi-stage
Bayesian inference~\cite{Roch:2025jpu}: neglecting correlations in the
likelihood. Cross-dataset correlations modify the information contributed
by different observables in a parameter-dependent manner, rather than
producing a uniform overcounting of information. Consequently, factorized
sequential updates do not generally reproduce joint inference when such
correlations are present. Since factorized updating is equivalent to using
a joint likelihood with the cross-dataset covariance omitted, the same
issue can arise in joint analyses that neglect these correlations.

To illustrate how cross-dataset correlations can be accounted for
in practice, consider a heavy-ion calibration in which a model has
already been constrained by a dataset $D_1$, and a new dataset $D_2$
becomes available. First, the measured data $d_1$, model predictions
$m_1(\theta)$, covariance $\Sigma_1$, and prior $p(\theta)$ are used to
obtain the posterior $p(\theta|D_1)$. Next, one determines the
predictions $m_2(\theta)$, covariance $\Sigma_2$, and cross-covariance
$\Sigma_{21}$ with the earlier dataset. For a Gaussian likelihood,
these inputs give the conditional mean $\mu_{2|1}(\theta)$ and covariance
$\Sigma_{2|1}$ through Eqs.~\eqref{eq:mean} and \eqref{eq:schur}.
The resulting conditional likelihood is
$p(D_2|D_1,\theta)=\mathcal{N}(d_2;\mu_{2|1}(\theta),\Sigma_{2|1})$,
which is used to update the previous posterior according to
Eq.~\eqref{eq:seq}. Importantly, the conditional mean depends on the
residual $d_1-m_1(\theta)$ of the earlier dataset. Thus, the previous
posterior alone is generally insufficient for this update; the earlier
measurements, model predictions, and cross-dataset covariance are also
needed.

This procedure is not restricted to particular observables or collision
conditions. The datasets may involve different centralities, energies,
or collision systems, or entirely different observable classes, including
soft- and hard-sector measurements, provided that a common model predicts
them and their joint likelihood can be specified. In practice, experimental
cross-covariances may arise from shared systematic uncertainties or
overlapping event samples, and can be assessed through experimental
uncertainty studies or joint resampling procedures. Correlated theoretical
or emulator uncertainties must also be characterized by the modeling
analysis. Importantly, shared sensitivity to the same physical parameters
does not by itself imply correlated uncertainties at fixed parameters.
When cross-covariances are unknown, varying plausible assumptions can
test the sensitivity of the inference, but cannot determine the exact
conditional likelihood.

Updating the physics model itself is different from incorporating a new
dataset. For example, if improved simulations change the predictions for
previously analyzed measurements, the likelihood for those measurements
must be updated, rather than multiplied by the old likelihood as if
the same data provided independent evidence. When the parameter space
and prior remain unchanged, the previous posterior can in principle be
reweighted by the ratio of the updated to the original likelihood,
provided that the existing samples adequately cover the new posterior.
More substantial changes in the parameterization, or insufficient
posterior coverage, may instead require a new calibration.

\section{Summary and conclusions}\label{sec:summary}

In this work, we have investigated sequential Bayesian inference in the presence
of correlated datasets within a controlled framework. While Bayesian updating
can always be formulated sequentially through the conditional likelihood,
practical implementations often rely on a factorized update that assumes
conditional independence. We have shown that this approximation is exact
under conditional independence, but generally leads to deviations from the
joint posterior when cross-dataset correlations are neglected.

Using a reduced model with a tunable covariance structure, we find that,
in our numerical examples with positive correlations, factorized sequential
updates depart increasingly from the joint result as the correlation strength
grows, while conditional updating reproduces the joint posterior exactly.
These deviations can be understood through an information decomposition that
separates the joint information into the contribution from the first dataset
and the conditional contribution from the second. Comparing the marginal and
conditional information reveals how neglecting correlations modifies the
constraints on different parameter combinations. These modifications reflect
the interplay between the covariance structure and parameter sensitivities,
providing a quantitative interpretation of the resulting posterior discrepancies.

More generally, factorized sequential updating corresponds to performing a
joint inference under an implicit assumption of conditional independence
between datasets given the model parameters.
When this assumption is violated, the resulting posterior can exhibit
distorted parameter constraints, including shifts in posterior estimates and
overestimated or underestimated uncertainties along different parameter
directions. Consistent sequential inference therefore requires a proper
treatment of correlations at the likelihood level within a common probabilistic
framework.

These considerations extend beyond explicitly sequential workflows.
In many analyses, posterior distributions obtained from previous datasets
are reused as priors when incorporating additional measurements. Such reuse
is fully consistent with Bayesian inference, but multiplying the previous
posterior by the marginal likelihood of new data implicitly assumes conditional
independence from the previously incorporated data. When this assumption
does not hold, the appropriate conditional likelihood is required to preserve
consistency with joint inference.

Although demonstrated in a reduced setting, the conclusions are broadly
applicable to multi-dataset Bayesian analyses.
The conditional-information diagnostics help characterize how statistical
dependence modifies constraints on different parameter combinations, while
exact conditional updating provides a consistent way to incorporate additional
datasets within a common probabilistic model.

\section*{Acknowledgements}

The author acknowledges valuable discussions with S. Jaiswal, P. M. Jacobs, and other members of the JETSCAPE Collaboration.
This work was supported in part by the U.S.
Department of Energy, Office of Science, Office of Nuclear Physics under
Grant No.~DE-AC02-05CH11231.

\section*{Declaration of generative AI and AI-assisted technologies}

During the preparation of this work the author used ChatGPT for grammar
refinement and clarity improvement. After using this tool/service, the author reviewed and edited the content as needed and takes full responsibility for the content of the published article.

\appendix
\section{Model construction and implementation}\label{app:model}

In this appendix, we provide the explicit construction of the reduced model used
in the main text. The purpose of this model is not to reproduce realistic
heavy-ion data, but to provide a controlled and transparent setting in which
different classes of observables probe partially overlapping aspects of a small
set of parameters. The guiding principle is to retain the minimal structure
necessary to capture multi-observable inference, while keeping the model simple
enough to allow direct evaluation of the likelihood without introducing
additional complications such as emulators or surrogate models.

The model depends on four parameters,
$\theta = (S,\, \varepsilon_2,\, \kappa_N,\, \kappa_v)$.
Here $S$ represents an overall initial scale (entropy-like quantity), and
$\varepsilon_2$ an eccentricity-like parameter controlling initial anisotropy.
The parameters $\kappa_N$ and $\kappa_v$ describe how efficiently these
quantities are converted into yield-like and flow-like observables,
respectively. In this way, $(S,\kappa_N)$ primarily control yields, while
$(\varepsilon_2,\kappa_v)$ primarily control flow, although this separation
is only approximate (see below).

We consider two groups of observables evaluated in three centrality bins
$i=1,2,3$, corresponding to representative central (0--10\%), mid-central (20--30\%), and peripheral (40--50\%)
collisions. The first group consists of yield-like observables $N_i$, and the
second of elliptic flow coefficients $v_{2,i}$, forming the data vector
$d = (N_1, N_2, N_3, v_{2,1}, v_{2,2}, v_{2,3})$.
To set realistic overall scales, we anchor the observables to fixed reference
values inspired by heavy-ion measurements \cite{ALICE2010, ALICE:2011ab}, using
$N^{\mathrm{ref}} = (1447.5, 649.0, 261.0)$ and
$v_2^{\mathrm{ref}} = (0.0359, 0.0831, 0.0994)$. These reference values encode
the dominant centrality dependence of the system, effectively capturing the
variation of entropy and eccentricity across centrality classes. All observables
are normalized at a reference parameter point
$\theta_\star = (S,\, \varepsilon_2,\, \kappa_N,\, \kappa_v)_\star=(1.0, 0.25, 1.0, 1.0)$,
so that the model reproduces the reference values at this point.

The observables are constructed multiplicatively as
\begin{equation}
N_i(\theta) =
N_i^{\mathrm{ref}}
\left(\frac{A_N(\theta)}{A_N(\theta_\star)}\right)
Y_i(\theta),
\quad
v_{2,i}(\theta) =
v_{2,i}^{\mathrm{ref}}
\left(\frac{A_v(\theta)}{A_v(\theta_\star)}\right)
V_i(\theta),
\end{equation}
which separates the dominant parameter dependence from subleading
bin-dependent effects.
The dominant structure is encoded in sector-wide amplitudes
$A_N(\theta) = \kappa_N S + \alpha_N \varepsilon_2$ and
$A_v(\theta) = \kappa_v \varepsilon_2 + \alpha_v S$,
with $\alpha_N = 0.12$ and $\alpha_v = 0.18$. These cross-coupling terms are
introduced to mimic the overlapping parameter sensitivities of
observables in realistic simulations. The choice $\alpha_v > \alpha_N$ ensures that the flow
observables retain a stronger sensitivity to the yield-side parameter $S$ than
vice versa, producing the asymmetric coupling structure discussed in the main text (see Fig.~\ref{fig2}).

To introduce distinct parameter sensitivities across centrality bins,
we include mild centrality-dependent shape factors. For the yields,
$Y_i(\theta) = 1 + a_i^{(S)}(S-S_\star) + a_i^{(\kappa_N)}(\kappa_N-\kappa_{N,\star})$, and for the
flow observables,
$V_i(\theta) = 1 + b_i^{(\varepsilon_2)}(\varepsilon_2-\varepsilon_{2,\star})
+ b_i^{(\kappa_v)}(\kappa_v-\kappa_{v,\star})$. The coefficients, $a_i$ and $b_i$, are chosen to be small, so that these terms act
as perturbations around the dominant amplitude dependence. Their role is to
introduce mild bin-to-bin variations and to avoid exact degeneracies, without
obscuring the main parameter structure.

Pseudo-data are generated from the reference point as
$d = m(\theta_\star) + \delta$, with Gaussian fluctuations
$\delta \sim \mathcal{N}(0,\Sigma)$. The covariance matrix is constructed in
block form to separate correlations within and between the two observable
groups.
Within each observable group, bin-to-bin correlations are modeled using a
short-range kernel \cite{JETSCAPE:2021ehl,Ehlers:2024miy}. For centrality bins $i,j=1,2,3$, we define a correlation
matrix $C_{ij} = \exp(-|i-j|/\ell)$ with correlation length $\ell = 1.0$, so that
neighboring bins are more strongly correlated than distant ones.
To retain a finite level of independent fluctuations, this correlated structure
is combined with an uncorrelated component. The resulting intra-group covariance
matrix for a given observable type $X \in \{N, v_2\}$ is constructed as
\begin{equation}
\Sigma_X^{ij}
=
\sigma_{X,i}\,\sigma_{X,j}
\left[
(1 - f_{\mathrm{ind}})\, C_{ij}
+ f_{\mathrm{ind}}\, \delta_{ij}
\right],
\end{equation}
where $f_{\mathrm{ind}} = 0.4$ controls the fraction of independent noise and
$\delta_{ij}$ is the Kronecker delta. This form ensures that correlations decay
with bin separation while preserving nonzero diagonal dominance.
The overall fluctuation scale is set by ALICE-inspired absolute uncertainties
for the selected Pb--Pb 2.76~TeV reference bins, multiplied by a uniform
rescaling factor of $1.7$. For the yield-like and flow-like observables, this
corresponds to
$\sigma_N^{\mathrm{obs}} = 1.7\,\sigma_N^{\mathrm{ref}}$, and
$\sigma_{v_2}^{\mathrm{obs}} = 1.7\,\sigma_{v_2}^{\mathrm{ref}}$,
where $\sigma^{\mathrm{ref}}$ denotes the baseline experimental uncertainties.
This choice provides a simplified heavy-ion-inspired noise model anchored to
realistic observable scales, while avoiding a detailed reproduction of the full
experimental covariance.

\begin{figure}[!t]
\centering
\includegraphics[width=0.7\linewidth]{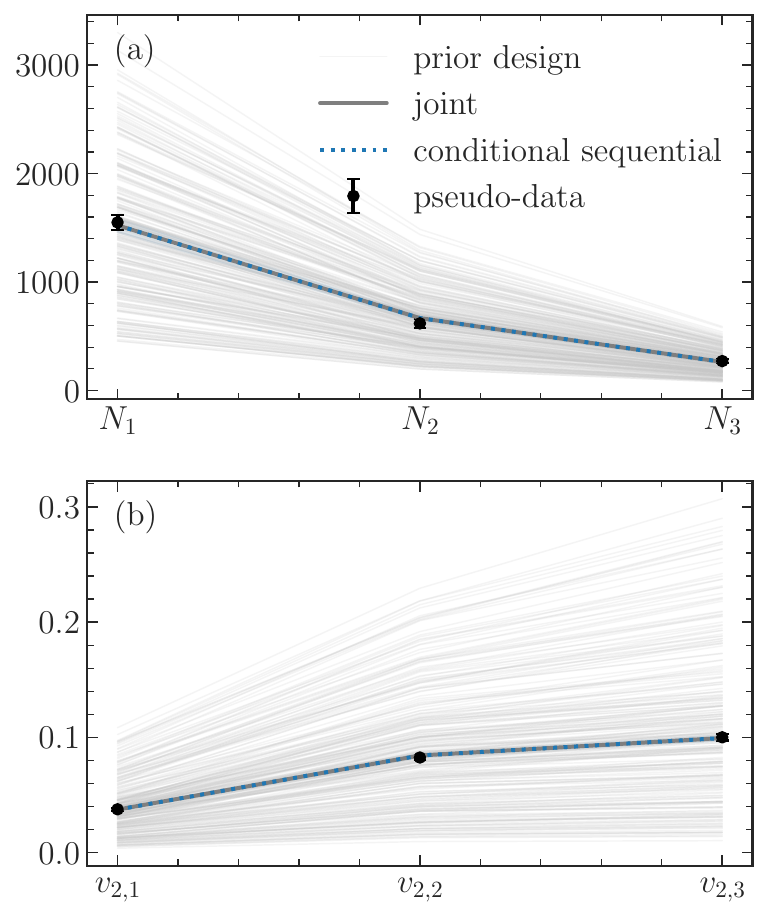}
\caption{
Model predictions for yield-like (a) and flow-like (b) observables at
$\rho=0.7$. Gray lines denote prior design points, markers show pseudo-data,
and curves correspond to joint and conditional sequential inference results.
}
\label{fig:a1}
\end{figure}

Cross-dataset correlations are introduced through the off-diagonal block
$\Sigma_{12}$ of the full covariance matrix \eqref{eq:covariance}. We construct it in factorized form as
\begin{equation}
\Sigma_{12}^{ij}
=
\rho\, \sigma_{N,i}\,\sigma_{v_2,j}\, C_{ij},
\end{equation}
where the same short-range kernel $C_{ij}$ is used to correlate nearby bins
across the two observable groups. The parameter $\rho$ controls the overall
strength of cross-dataset correlations, with $\rho=0$ corresponding to
statistically independent datasets and increasing positive values introducing
stronger correlations.
Negative values of $\rho$ describe anti-correlations and are treated by the
same construction, provided that the full covariance matrix remains positive
definite. The numerical results in this work use $0\leq\rho\leq0.7$, within
the positive-definite range of the specified covariance.
The full covariance matrix then takes the block form
$\Sigma = \begin{pmatrix} \Sigma_1 & \Sigma_{12} \\ \Sigma_{21} & \Sigma_2 \end{pmatrix}$,
with $\Sigma_1 \equiv \Sigma_N$ and $\Sigma_2 \equiv \Sigma_{v_2}$ (see Fig.~\ref{fig1}).

Finally, we adopt a uniform prior over the parameter ranges
$S \in [0.5,1.5]$, $\varepsilon_2 \in [0.05,0.5]$,
$\kappa_N \in [0.5,1.5]$, and $\kappa_v \in [0.5,1.5]$.
These ranges are chosen to be broad enough to encompass the relevant parameter
space without imposing strong prior constraints. Parameter points leading to
nonphysical (negative) observables are excluded by assigning them vanishing
likelihood.
Figure~\ref{fig:a1} illustrates the resulting structure of the model and data.
The prior design points span a broad region of observable space, while the
posterior predictions from joint and conditional sequential inference
closely reproduce the pseudo-data. Here, the posterior predictions correspond to the
posterior predictive mean, obtained by averaging model predictions over the
full posterior distribution. This demonstrates that the model retains sufficient
flexibility to describe the data while remaining simple enough to isolate the
effects of correlations.

\section{Matrix structure of correlation effects}
\label{app:matrix}

To clarify how correlations propagate from the data level to parameter
inference, we examine the Gaussian likelihood in block form. The goal is to make
explicit how the covariance structure modifies the information carried by
different datasets, and how this modification is encoded in both the likelihood
and its parameter-space representation.

We introduce residuals $r_1 = d_1 - m_1(\theta)$ and
$r_2 = d_2 - m_2(\theta)$, and write the covariance matrix as
$$\Sigma = \begin{pmatrix}
\Sigma_1 & \Sigma_{12} \\
\Sigma_{21} & \Sigma_2
\end{pmatrix},$$
where $\Sigma_{12}$ encodes cross-dataset correlations. The joint Gaussian
likelihood can be written as
\begin{equation}
\mathcal{L}(\theta) = p(D_1,D_2|\theta)
\propto
\exp\left[-\frac12\, Q_{\rm joint}\right],
\end{equation}
with quadratic form
\begin{equation}
Q_{\rm joint}
=
\begin{pmatrix}
r_1 \\
r_2
\end{pmatrix}^T
\Sigma^{-1}
\begin{pmatrix}
r_1 \\
r_2
\end{pmatrix}.
\end{equation}
The quadratic form measures the mismatch between data and model predictions,
weighted by the inverse covariance. Correlations between datasets therefore
enter through the off-diagonal blocks of $\Sigma^{-1}$, which couple the
residuals $r_1$ and $r_2$.

Using the block inverse of $\Sigma$, the quadratic form can be written as
\begin{equation}
Q_{\rm joint}
=
r_1^T\Sigma_1^{-1}r_1
+
\left(
r_2 - \Sigma_{21}\Sigma_1^{-1}r_1
\right)^T
\Sigma_{2|1}^{-1}
\left(
r_2 - \Sigma_{21}\Sigma_1^{-1}r_1
\right),
\end{equation}
where $\Sigma_{2|1} = \Sigma_2 - \Sigma_{21}\Sigma_1^{-1}\Sigma_{12}$ is the
Schur-complement covariance, representing the uncertainty in $D_2$ that
remains after accounting for correlations with $D_1$.
This identity makes explicit that the joint likelihood factorizes exactly as
$p(D_1,D_2|\theta) = p(D_1|\theta)\, p(D_2|D_1,\theta)$. The second term can be
written in terms of a conditional residual
\begin{equation}
r_{2|1}
=
r_2 - \Sigma_{21}\Sigma_1^{-1} r_1
=
d_2 - \mu_{2|1}(\theta),
\end{equation}
where the conditional mean is given by
\begin{equation}
\mu_{2|1}(\theta)
=
m_2(\theta)
+
\Sigma_{21}\Sigma_1^{-1}
\bigl(d_1 - m_1(\theta)\bigr).
\end{equation}
In this form, the second term describes fluctuations of $D_2$ around the
conditional mean $\mu_{2|1}(\theta)$, accounting for the part of the
statistical fluctuations in $D_2$ that can be predicted from $D_1$ at fixed
model parameters.

To connect this structure to parameter inference, we consider the expected
local curvature of the Gaussian log-likelihood for parameter-independent
covariance. Writing $\delta\theta = \theta-\theta_\star$ and taking the
expectation over pseudo-data generated at $\theta_\star$, the expected
log-likelihood has the expansion
\begin{equation}
\mathbb{E}_{d|\theta_\star}\bigl[\log \mathcal{L}(\theta)\bigr]
\simeq
-\frac12\, \delta\theta^T F_{\rm joint}\, \delta\theta
+ \mathrm{const},
\end{equation}
where the Fisher information matrix is
\begin{equation}
F_{\rm joint} = J^T \Sigma^{-1} J
\end{equation}
with $J = \partial m/\partial\theta$ evaluated at $\theta_\star$.
This is the expected curvature, which can differ from the observed curvature
for a particular pseudo-data realization. The Gaussian likelihood decomposition
therefore induces a corresponding decomposition of Fisher information.

Differentiating the conditional mean yields the modified Jacobian
\begin{equation}
J_{2|1}
=
\frac{\partial \mu_{2|1}(\theta)}{\partial \theta}
=
J_2 - \Sigma_{21}\Sigma_1^{-1} J_1,
\end{equation}
where $J_i = \partial m_i/\partial\theta$. The corresponding Fisher contribution
is
\begin{equation}
F_{2|1} = J_{2|1}^T \Sigma_{2|1}^{-1} J_{2|1},
\end{equation}
while
\begin{equation}
F_1 = J_1^T \Sigma_1^{-1} J_1.
\end{equation}
The joint Fisher matrix can therefore be written as
\begin{equation}
F_{\rm joint} = F_1 + F_{2|1}.
\end{equation}
This decomposition provides the parameter-space counterpart of the likelihood
factorization above: the total information is the sum of the contribution from
$D_1$ and the conditional information contribution from $D_2$.
In this representation, the effects
of cross-dataset correlations do not appear as explicit cross terms, but are
absorbed into the modified covariance $\Sigma_{2|1}$ and response $J_{2|1}$.

\begin{figure}
    \centering
    \includegraphics[width=\linewidth]{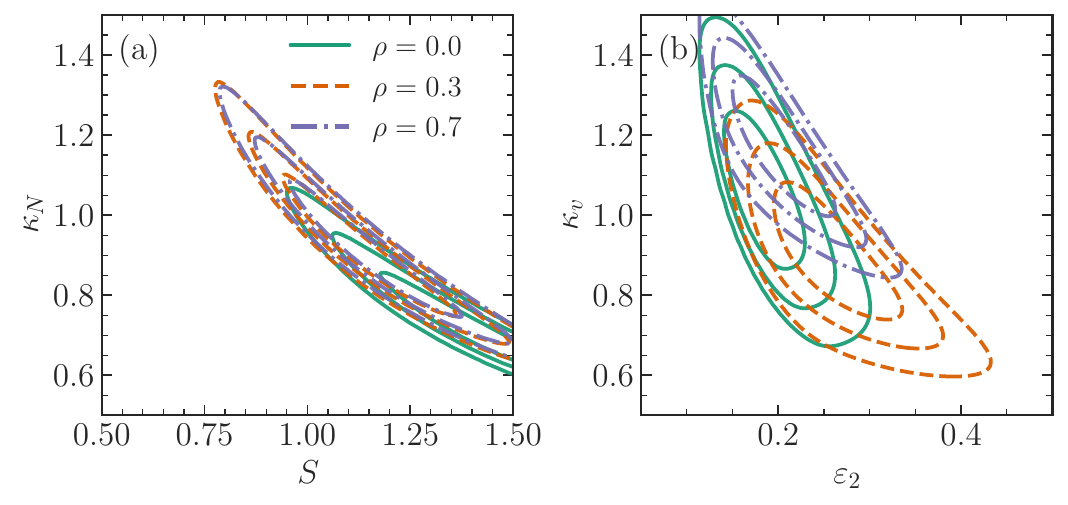}
\caption{
Joint posterior distributions for three representative values of the
cross-dataset correlation strength $\rho$. The left panel shows the
$(S,\kappa_N)$ plane and the right panel the $(\varepsilon_2,\kappa_v)$ plane.
Contours correspond to $\rho=0.0$ (solid), $\rho=0.3$ (dashed), and
$\rho=0.7$ (dash-dotted). All posteriors are obtained using the full joint
likelihood.
}
    \label{fig:a2}
\end{figure}

From a complementary perspective, the same mechanism can be understood in terms
of the shape of the likelihood over parameter space. Writing the inverse
covariance in block form, the off-diagonal blocks of $\Sigma^{-1}$ couple the
residuals of the two datasets. The quadratic form contains a cross term
$2r_1^T(\Sigma^{-1})_{12}r_2$, where $(\Sigma^{-1})_{12}$ denotes the
upper-right block of the inverse covariance. This term encodes the effect of
cross-dataset correlations. The posterior is therefore altered through a
reshaping of the likelihood landscape, with the magnitude and direction of the
effect determined by how the two datasets respond jointly to variations in
$\theta$. The impact along a given parameter direction depends on how its
changes in the residuals of the two datasets project onto this correlated
covariance structure.

This behavior is visible directly in Fig.~\ref{fig:a2}. As $\rho$ increases,
the posterior contours change, reflecting both the covariance structure and
the associated pseudo-data realizations in this numerical study.
The changes are more pronounced in the $(\varepsilon_2,\kappa_v)$ plane than in the
$(S,\kappa_N)$ plane, consistent with the asymmetric sensitivity structure
discussed in the main text. In this way, the off-diagonal structure of
$\Sigma^{-1}$ determines how correlations are projected into parameter space
locally, while the induced reshaping of the likelihood governs how these effects
accumulate to produce the global posterior distortions discussed in the main
text.

\bibliographystyle{elsarticle-num} 
\bibliography{refs}

\end{document}